\documentclass[10pt]{iopart}
\usepackage{iopams}

\begin{document}

\noindent Version 27Nov07:12:45, ultima revision MS

\title{Classical and quantum integrability in 3D systems.}

\author{M. Gadella$^1$, J. Negro$^1$,  G.P. Pronko$^{2,3}$, M.
Santander$^1$}

\address{$^1$Departamento de F\'isica Te\'orica. Facultad de Ciencias\\
47011 Valladolid, Spain.
\\
$^2$Institute for High Energy Physics , Protvino, Moscow reg.,Russia
\\
$^3$Institute of Nuclear Physics, National Research Center
``Demokritos", Athens, Greece.}

\ead{manuelgadella@yahoo.com.ar, jnegro@fta.uva.es, pronko@ihep.ru,
msn@fta.uva.es}

\begin{abstract}

In this contribution, we discuss three situations in which complete
integrability of a three dimensional classical system and its
quantum version can be achieved under some conditions. The former is
a system with axial symmetry. In the second, we discuss a three
dimensional system without spatial symmetry which admits separation of
variables if we use ellipsoidal coordinates. In both cases, and as a
condition for integrability, certain conditions arise in the
integrals of motion. Finally, we study integrability in the three
dimensional sphere and a particular case associated with the Kepler
problem in $S^3$.

\end{abstract}

\section{Introduction.}

For a system whose configuration space has dimension $n$, integrability in the sense of Liouville-Arnold requires the existence of a number $n$ of constants of motion which should be in involution (the Hamiltonian itself is one of them). Integrable systems are rare, yet many systems which are important from a physical standpoint turn out to be integrable, with the outstanding examples of the harmonic oscillator and the Kepler problem, which in fact are superintegrable (they have more than $n$ functionally independent constants of motion, albeit not all of them are in involution). 

A type of integrability appears to be particularly relevant: the {\em quadratic integrability}. 
A standard natural Hamiltonian has a kinetic part which is {\it quadratic} in the momenta. If all $n-1$ additional constants of motion have a similar kind of structure, we will speak of quadratic integrability (understanding quadratic as `at most quadratic', i.e., allowing possibly for constants of motion which are linear in the momenta).

For a 2D system in any constant curvature configuration space, quadratic integrability means the existence of a single constant of motion quadratic in the momenta; this case is simpler as there is no any extra condition ensuring the additional constants are in involution \cite{BaHeSaSa03,SaSa05}. For any space of constant curvature, integrable systems have a Hamiltonian which is  separable in confocal (or general elliptic) coordinates in that space \cite{Benen97, Cram03}. These coordinates include generic ones (as the elliptic coordinates in the sphere $S^2$ or the Euclidean plane $E^2$) as well as all its possible degenerations or limiting cases. This also holds in the 3D case \cite{HoMcLSm05, Ev90}. It is generally believed that a similar connection among quadratic integrability and the $n$D version of confocal coordinates does somehow hold, but to the best of our knowledge a proof is not available in the literature. 

In this work we explore some aspects of quadratic integrability for a system in 3D configuration spaces $E^3$ or $S^3$. The results should be considered as a stage towards studying  how precisely integrable systems in Euclidean or in any constant curvature 3D space are related to separable systems in the generic version of ellipsoidal coordinates, associated to confocal quadrics \cite{Ka99,Ka02}. We discuss mainly the Euclidean case, but extension of these results to spaces with any constant curvature seems to be possible, and as a hint in that direction, we discuss the integrability with axial symmetry and the Kepler problem in a space with positive constant curvature $S^3$ (see also \cite{HeBa06}).

\section{3D systems with axial symmetry.}

In this first part, we consider a three dimensional system in
Euclidean space with axial symmetry around an axis given by the unit
vector ${\bf n}=(n_1,n_2,n_3)$. The classical analysis of this
system makes use of canonical coordinates $({\bf p},{\bf
x})=(p_i,x_i)$, $i=1,2,3$ and Poisson brackets
$\{A,B\}=\frac{\partial A}{\partial p_k}\frac{\partial B}{\partial
x_k}-\frac{\partial A}{\partial x_k}\frac{\partial B}{\partial
p_k}$. We want to characterize those systems with axial symmetry
which are completely integrable. For these systems dynamics is described by a Hamiltonian of the form
\begin{equation}\label{1}
    H= \frac{{\bf p}^2}{2m} + U({\bf x})\,,
\end{equation}
where $U(\bf x)$ is a time independent potential. Axial symmetry requires that the angular momentum $L$ around the symmetry axis
\begin{equation}\label{2}
L = {\bf n}\cdot\left({\bf x}\times {\bf p}\right)
\end{equation}
be a constant of motion, i.e., $\{H,L\}=0$. Complete quadratic integrability
requires  the existence of a third independent constant of motion $H_1$,
which is in involution with $H$ and $L$, i.e.,
$\{H,H_1\}=\{L,H_1\}=0$ and has a `Hamiltonian form' in general position dependent yet quadratic in the momenta, 
plus a time independent `potential', with form
\begin{equation}\label{3}
 H_1=\frac{1}{2m}\,p_i\,g^{ik}({\bf x})\,p_k + \Phi({\bf x})\,.
\end{equation}
The tensor $g^{ik}({\bf x})$, which determines the `kinetic' part of
the constant of motion $H_1$ has to be a Killing tensor for the Euclidean metric. Its determination can be done by assuming that the `kinetic' terms in (\ref{1}) and (\ref{3}) commute among themselves
and also with the angular momentum $L$, with respect to the Poisson
bracket. If we choose the angular momentum direction as the $z$ axis, then the most general expression for $\,p_i\,g^{ik}({\bf x})\,p_k$ is
\cite{Wi67}:
\begin{equation}\label{1a}
    \,p_i\,g^{ik}({\bf x})\,p_k = L_1^2 + L_2^2 + \alpha p_3^2 +\beta (p_1L_2 - p_2L_1) + \gamma p_3L_3
\end{equation}
to which of course any linear combination of the quadratic Casimirs of the Euclidean algebra, $p_1^2 + p_2^2 + p_3^2$ and $p_1L_1 + p_2L_2 + p_3L_3$ can be added. Here we will discuss mainly the family with $\alpha<0, \beta=0, \gamma=0$, for which 
\begin{equation}\label{1b}
   \,p_i\,g^{ik}({\bf x})\,p_k = L_1^2 + L_2^2 -a^2 p_3^2
\end{equation}
with $a$ a positive constant. If now the angular momentum direction is along a general unit vector ${\bf n}=(n_1,n_2,n_3)$, by  direct computation we get for the Killing tensor $g^{ik}({\bf x})$ the expression
\begin{equation}\label{4}
g^{ik}({\bf x})= \delta^{ik} ({\bf x}\cdot{ \bf n})^2 - {\bf x}\cdot
{\bf n} (x_i n_k + x_k n_i) + ({\bf x}^2- a^2)n_i n_k\,,
\end{equation}
For more details, see \cite{GaNePr07}. The commutation
of $L$ with $H$ and $H_1$ restricts the form of the functions $U$ and $\Phi$, collectively called `potential' terms to:

\begin{equation}\label{5}
U({\bf x}) = U({\bf x}^2,({\bf x}\cdot{ \bf n})^2),\qquad \Phi({\bf
x}) = \Phi({\bf x}^2,({\bf x}\cdot{ \bf n})^2)
\end{equation}
(note that the condition of axial symmetry around the $\bf n$ axis implies that $U$ and $\Phi$ cannot depend on the azimuthal angle $\varphi$) and the commutation of $H$ and $H_1$ leads to the equations:

\begin{equation}\label{6}
\partial_{i} \Phi({\bf x}) =
g^{ik}({\bf x})\partial_{k} U({\bf x})\,,\qquad i,k=1,2,3\,,
\end{equation}
where we always assume summation over repeated indices. In order to
obtain the solutions of (\ref{6}), we diagonalize the matrix
with components $g^{ik}({\bf x})$. Its eigenvalues $\lambda({\bf
x})$ and eigenvectors $A({\bf x})$ are obtained from the matrix
equation

\begin{equation}\label{7}
    \left(g^{ik}({\bf x})- \lambda({\bf x}) \delta^{ik}\right)
A_k({\bf x})=0\,.
\end{equation}

The eigenvalue equation (\ref{7}) has the following solutions:

\begin{equation}\label{8}
    \lambda_{\pm}({\bf x}) = \frac{{\bf
x}^2-a^2}{2}\pm\sqrt{\left(\frac{{\bf x}^2-a^2}{2}\right)^2+ a^2\,
{\bf x \bf n}^2},\qquad \lambda_0({\bf x})= {\bf x \bf n}^2\,,
\end{equation}
with corresponding eigenvectors given by

\begin{equation}\label{9}
A^\pm_i({\bf x})=\partial_i \lambda_\mp({\bf x})\,,\qquad A^0_i({\bf
x})= \frac{{\bf n}\times{\bf x}}{{\bf x}^2-{\bf x \bf n} ^2}=
\partial_{i}\varphi({\bf x})\,,
\end{equation}
where the index $i=1,2,3$ labels the component of the corresponding
vector field and $\varphi$ is the azimuthal angle around the ${\bf
n}$-axis. These vectors form a new coordinate basis, $\{
\mathbf{\nabla}\lambda_-,\mathbf{ \nabla}\lambda_+,\mathbf{
\nabla}\varphi\}$ referred to which the matrix $g^{ik}({\bf
x})$ is diagonal. Taking into account that, due to the geometric
symmetry, $U$ and $\Phi$ cannot depend on $\varphi$, (\ref{6})
becomes

\begin{equation}\label{10}
\partial_i\lambda_+\,\partial_+\Phi +
 \partial_i\lambda_-\,\partial_-\Phi=
g^{ik}\left(\partial_k\lambda_+\,\partial_+U +
\partial_k\lambda_-\,\partial_-U\right)\,,
\end{equation}
where $\partial_\pm$ stand for
$\frac{\partial}{\partial\lambda_\pm}$. In the basis $\{\mathbf{\nabla}\lambda_-,\mathbf{ \nabla}\lambda_+,\mathbf{
\nabla}\varphi\}$ that
diagonalizes the matrix $g^{ik}({\bf x})$, equation (\ref{10}) decouples in

\begin{equation}\label{11}
    \partial_+ \Phi = \lambda_- \partial_+U\,,\qquad \partial_- \Phi =
 \lambda_- \partial_+U
    \,.
\end{equation}
This means that

\begin{equation}\label{12}
\Phi - \lambda_+ U = -f(\lambda_+),\qquad   \Phi - \lambda_- U =
-g(\lambda_-)\,,
\end{equation}
where $f(\lambda_+)$ and $g(\lambda_-)$ are arbitrary functions on
their respective variables. These equations give an expression for
the `potentials' as follows:

\begin{equation}\label{13}
U=\frac{f(\lambda_+)-g(\lambda_-)}{\lambda_+-\lambda_-},\qquad
\Phi=\frac{\lambda_-}{\lambda_+-\lambda_-}\,f(\lambda_+) -
\frac{\lambda_+}{\lambda_+-\lambda_-}\,g(\lambda_-)\,.
\end{equation}
These are the most general expressions for the potentials $U({\bf
x})$ and $\Phi({\bf x})$ compatible with our choice for $H_1$ and with the requirement of $H, H_1, L$ being in involution. 

A comment is here in order: the presence of a constant of motion which is first-order in the momenta, i.e. is a Noether constant, means that the system is invariant under rotations around the axis ${\bf n}$. Then this situation reduces to one that is essentially two-dimensional, with a single azimutal coordinate $\varphi$ added to the ordinary 2D elliptic coordinates in a fixed plane containing the axis ${\bf n}$. In the 3D Euclidean space these coordinates are the spheroidal (oblate) cordinates \cite{MiPaZa81}, which play the role of 3D separation coordinates for the system, as we will show explicitly in the next section.  

\subsection{Integration.}

The integration of the classical system with axial symmetry with
respect to the $\bf n$ axis is based in the following idea: If
$e_i=h({\bf p},{\bf x})$, $i=1,2,3$ are three independent functions
with $\{e_i,e_j\}=0$, then, there exists a function $F({\bf x},{\bf
e})$ with ${\bf e}=(e_1,e_2,e_3)$ and

\begin{equation}\label{14}
p_k=\frac{\partial F({\bf x},{\bf e})}{\partial x_k}\,.
\end{equation}
The function $F({\bf x},{\bf e})$ is the characteristic function in
the Hamilton-Jacobi approach.

In our case, the chain rule and (\ref{14}) gives the following
expression for the momentum ${\bf p}=(p_1,p_2,p_3)$:

\begin{equation}\label{15}
\fl
{\bf p} = \mathbf{\nabla}\lambda_+\,
\partial_+F(\lambda_+,\lambda_-,\varphi)+ \mathbf{\nabla}\lambda_-\,
\partial_-F(\lambda_+,\lambda_-,\varphi)+ \frac{{\bf n}\times{\bf
x}}{{\bf x}^2-({{\bf x}\cdot{\bf n}})^2}\,
\partial_\varphi F(\lambda_+,\lambda_-,\varphi)\,.
\end{equation}
Then, the kinetic part of $H$ can be written in the following form:

\begin{equation}\label{16}
\fl
{\bf p}^2 = (\mathbf{\nabla}\lambda_+)^2 (\partial_+ F)^2+
(\mathbf{\nabla}\lambda_-)^2 (\partial_- F)^2+ \frac{\ell^2}{{\bf
x}^2-({{\bf x}\cdot{\bf n}})^2}= 2m\left(E-U\right)\,,
\end{equation}
where $\ell$ is the value of the integral of motion $L$
corresponding to the angular momentum around ${\bf n}$, i.e.,

\begin{equation}\label{17}
\left({\bf n}\times{\bf x}\right)\cdot {\bf p} = {\bf n}\cdot {\bf
L} = \ell=\partial_\varphi F(\lambda_+,\lambda_-,\varphi)
\end{equation}
and $E$ is a given value of the constant of motion $H$.

For the kinetic term of $H_1$, we obtain a similar expression:

\begin{equation}\label{18}
\fl
p_ig^{ik}p_k = \lambda_-(\nabla \lambda_+)^2 (\partial_+ F)^2+
\lambda_+(\nabla\lambda_-)^2 (\partial_- F)^2+ \frac{({{\bf
x}\cdot{\bf n}})^2\,\ell^2}{{\bf x}^2-({{\bf x}\cdot{\bf n}})^2}=2m(
E_1 - \Phi)\,,
\end{equation}
where $E_1$ is a constant value of $H_1$.

With some calculations \cite{GaNePr07}, we obtain the partial derivatives
of the function $F(\lambda_+,\lambda_-,\varphi)$ in terms of its
arguments:
\begin{eqnarray}
\fl
\displaystyle
 (\partial_{+}F)^2
=
\frac{1}{4\lambda_{+}(\lambda_{+}+a^2)}\left\{2m\left(\lambda_{+}(E-U)-(
E_1-\Phi)\right)
-\frac{ \lambda_{+}l^2}{\lambda_{+}+a^2}\right\}\,, \label{19}\\[2ex]
\fl
\displaystyle
 (\partial_{-}F)^2
=
\frac{1}{4\lambda_{-}(\lambda_{-}+a^2)}\left\{2m\left(\lambda_{-}(E-U)-(
E_1-\Phi)\right) -\frac{ \lambda_{-}l^2}{\lambda_{-} +a^2}\right\}
\,. \label{20}
\end{eqnarray}

Due to (\ref{12}), (\ref{19}) does not depend on $\lambda_-$ and
(\ref{20}) does not depend on $\lambda_+$. From this simple idea, we
conclude that the function $F(\lambda_+,\lambda_-,\varphi)$ is of
the form:

\begin{equation}\label{21}
F(\lambda_+,\lambda_-,\varphi)=A(\lambda_+)+B(\lambda_-)+\ell\varphi\,,
\end{equation}
where $A(\cdot)$ and $B(\cdot)$ are functions of one variable only.
Thus, we have separated variables in the Hamilton-Jacobi generating
function. This permits us to obtain the equations of motion in terms
of the variables $(\lambda_+,\lambda_-,\varphi)$. The final result
is \cite{GaNePr07}
\begin{eqnarray}
\displaystyle &\frac{\dot
\lambda_{+}}{\lambda_{+}\partial_{+}A(\lambda_{+})} +\frac{\dot
\lambda_{-}}{\lambda_{-}\partial_{-}B(\lambda_{-})}
=\frac{4}{m} \nonumber\\[2ex]
\displaystyle &\frac{\dot
\lambda_{+}}{\lambda_{+}(\lambda_{+}+a^2)\partial_{+}A(\lambda_{+})}
+\frac{\dot
\lambda_{-}}{\lambda_{-}\partial_{-}(\lambda_{-}+a^2)B(\lambda_{-})}=0
\nonumber\\[2ex]
\displaystyle &-\frac{la^2}{4}\left(\frac{\dot
\lambda_{+}}{\lambda_{+}(\lambda_{+}+a^2)^2\partial_{+}A(\lambda_{+})}
+\frac{\dot
\lambda_{-}}{\lambda_{-}\partial_{-}(\lambda_{-}+a^2)^2B(\lambda_{-})}\right)=\dot
\varphi\,, \label{22}
\end{eqnarray}
where the upper dot represents the derivative with respect time.
These equations are at least formally integrable.

\subsection{Quantum case.}

Canonical quantization of the functions $H,L,H_1$ give respective
operators that we represent with the same symbols. Then, complete
integrability means that

\begin{equation}\label{23}
    [H,L]=[H_1,L]=[H,H_1]=0\,,
\end{equation}
where $[A,B]=AB-BA$ denotes the commutator of the operators $A$ and
$B$. Equation (\ref{23}) implies the existence of simultaneous
eigenfunctions $\psi({\bf x})$ of these three operators:

\begin{equation}\label{24}
(H-E)\psi = (H_1-  E_1)\psi = (L - \ell)\psi=0\,,
\end{equation}
or equivalently,
\begin{eqnarray}
-\Delta \psi({\bf x}) = 2m(E- U({\bf x}))\psi({\bf x})
\label{25}\\[2ex]
 -\Delta_1 \psi({\bf x}) =
2m( E_1- \Phi({\bf x}))\psi({\bf x})\label{26} \\[2ex]
-i\partial_{\varphi} \psi({\bf x}) = \ell\, \psi({\bf
x})\,,\label{27}
\end{eqnarray}
where,

\begin{equation}\label{28}
\Delta = \partial_k\partial_k\,, \qquad  \Delta_1 = \partial_j
g^{jk}({\bf x}) \partial_k\,.
\end{equation}

We can express (\ref{25}) and  (\ref{26}) in terms of the variables
$\lambda_\pm$. Using (\ref{27}) we obtain respectively,
\begin{eqnarray}
\fl
-\left[4\lambda_{+}(\lambda_{+}{+}a^2)\psi_{++}
+2(a^2{+}3\lambda_{+})\psi_{+} - \frac{l^2 \lambda_{+} }{\lambda_{+}
{+}a^2}\psi \right]
= 2m \left[(E{-}U)\lambda_{+} {-} (
E_1 {-}\Phi)\right]\psi\, ,\label{29}\\[2ex]
\fl
-\left[4\lambda_{-}(\lambda_{-}{+}a^2)\psi_{--}
+2(a^2{+}3\lambda_{-})\psi_{-} - \frac{l^2 \lambda_{-} }{\lambda_{-}
{+}a^2}\psi \right]
 = 2m \left[(E{-}U)\lambda_{-} {-}
(E_1{-}\Phi)\right]\psi\,. \label{30}
\end{eqnarray}
Note that (\ref{29}) and (\ref{30}) depend only on $\lambda_+$ and
$\lambda_-$ respectively. Then, the wave function $\psi({\bf x})$
can be factorized as

\begin{equation}\label{31}
\psi({\bf x})=\psi^+(\lambda_+)\,\psi^-(\lambda_-)\,e^{i\ell\varphi}
\end{equation}
so that (\ref{28}), (\ref{29}) and (\ref{27}) can be written as
equations depending solely on the variables
$(\lambda_+,\lambda_-,\varphi)$ respectively. Once we have solved
equations (\ref{28}) and (\ref{29}), we have solved the problem of
finding solutions of (\ref{24}) or, equivalently,
(\ref{25}-\ref{27}). In order to finding solutions, we have to give
explicit expressions for the potentials $U$ and $\Phi$. These
expressions must be obtained from (\ref{13}) and the form of the
functions $f(\lambda_+)$ and $g(\lambda_-)$.

In order to illustrate the search for solutions of (\ref{29}) and
(\ref{30}), let us propose a simple although nontrivial choice of
$f(\lambda_+)$ and $g(\lambda_-)$ so that (\ref{28}) and (\ref{29})
are exactly solvable. Here, we propose

\begin{equation}\label{32}
f(\lambda_+) = \Phi-\lambda_+U = 0\,,\qquad g(\lambda_-) =
\Phi-\lambda_-U = - Q (\lambda_-+a^2)\,,
\end{equation}
where $Q$ is a constant. This choice give the following form for the
potentials:

\begin{equation}\label{33}
U({\bf x}) = -Q\, \frac{\lambda_-+a^2}{\lambda_+-\lambda_-}\,,\qquad
 \Phi({\bf  x}) = - Q\,
\frac{\lambda_+(\lambda_-+a^2)}{\lambda_+-\lambda_-}\,.
\end{equation}

The steps to solve (\ref{28}) and (\ref{29}) with these potentials
are the following: i) From (\ref{8}), we conclude that $\lambda_+$
is always positive meanwhile the values of $\lambda_-$ lie on the
interval $[-a^2,0]$. This suggest the following change of variables:
$\lambda_+ = a^2 \sinh^2\alpha$ and $\lambda_- = -a^2 \sin^2 \beta$.
The new coordinates $(\alpha,\beta,\varphi)$ are known as the {\it
oblate} spherical coordinates. This change of variables transforms
(\ref{28}) and (\ref{29}) into two new wave functions \cite{GaNePr07}.
ii) The new change of variables $t:=\sinh\alpha$ and $u:=\sin\beta$
and the introduction of the new parameters, ${\cal E}:=2ma^2E,
q:=2ma^2Q, {\cal E}_1:=\ell^2+2mE_1, G:={\cal E}_1+{\cal E}$ and
$q'=q-{\cal E}$, with $t=i\alpha$ transform equations (\ref{28}) and
(\ref{29}) respectively into
\begin{eqnarray}
\fl
 (1-\alpha^2) \frac{d^2\psi^+(\alpha)}{d \alpha^2} -
2\alpha\,\frac{d\psi^+(\alpha)}{d \alpha}
+\left\{G-\frac{\ell^2}{1-\alpha^2}-{\cal
E}(1-\alpha^2)\,\right\}\psi^+(\alpha)\,, \label{34}\\[2ex]
\fl
(1-u^2) \frac{d^2\psi^-(u)}{d u^2} - 2u\,\frac{d\psi^-(u)}{d u}
+\left\{ G- \frac{\ell^2}{1-u^2}  +q'(1-u^2)\, \right\}
\psi^-(u)=0\,.\label{35}
\end{eqnarray}

Equations (\ref{34}-\ref{35}) are spheroidal equations, a type
of second order differential equation which has been studied
\cite{MeScWoBKMaFu}. A brief discussion of the solutions for
(\ref{34}-\ref{35}) and therefore for (\ref{29}-\ref{30}) can be
found in \cite{GaNePr07}.

\section{General 3D integrable systems.}

In this section, we intend to generalize the previous discussion to
the case in which no (Noether) symmetry is present. As in the previous case,
we shall discuss the classical point of view first and then its
quantum counterpart. The key of the solution to this general case
will be the choice of a proper system of coordinates, as we shall
see. This justifies the next subsection.

\subsection{The ellipsoidal coordinates.}

The generic ellipsoidal coordinate system in the Euclidean three
dimensional space is determined as follows. Fix three positive
numbers $a$, $b$ and $c$ with $a>b>c$, and consider the one-parameter family of quadrics with equation
\begin{equation}\label{370}
 \frac{x^2}{a^2+\xi}+ \frac{y^2}{b^2+\xi}+
    \frac{z^2}{c^2+\xi}=1\,,
\end{equation}
Through any given point $(x,y,z)$ in Euclidean 3D space, there passes precisely three such quadrics, corresponding to the values $\lambda, \mu, \nu$ of the parameter $\xi$ which lie in the intervals
\begin{equation}\label{36}
-a^2< \nu < -b^2 < \mu < -c^2 < \lambda \, .
\end{equation}
One quadric is an ellipsoid (because $a^2+\lambda>b^2+\lambda>c^2+\lambda>0$)
\begin{equation}\label{37}
 \frac{x^2}{a^2+\lambda}+ \frac{y^2}{b^2+\lambda}+
    \frac{z^2}{c^2+\lambda}=1\, ,
\end{equation}
other is an one sheeted hyperboloid (because $a^2+\mu>b^2+\mu>0>c^2+\mu$)

\begin{equation}\label{38}
\frac{x^2}{a^2+\mu}+ \frac{y^2}{b^2+\mu}+
    \frac{z^2}{c^2+\mu}=1\,,
\end{equation}
and the last is a two sheeted hyperboloid (because $a^2+\nu>0>b^2+\nu>c^2+\nu$)

\begin{equation}\label{39}
\frac{x^2}{a^2+\nu}+ \frac{y^2}{b^2+\nu}+
    \frac{z^2}{c^2+\nu}=1\,.
\end{equation}

Solving (\ref{37}-\ref{39}) for $x^2$, $y^2$ and $z^2$, we obtain the parametrization of 3D Euclidean space in terms of ellipsoidal coordinates, $\lambda$, $\mu$ and $\nu$:
\begin{eqnarray}
x^2=\frac{(\lambda+a^2)(\mu+a^2)(\nu+a^2)}{(b^2-a^2)(c^2-a^2)}\,,
 \label{40}
\\[2ex]
 y^2=\frac{(\lambda+b^2)(\mu+b^2)(\nu+b^2)}{(c^2-b^2)(a^2-b^2)}\,,
 \label{41}
 \\[2ex]
z^2=\frac{(\lambda+c^2)(\mu+c^2)(\nu+c^2)}{(a^2-c^2)(b^2-c^2)}\,.\label{42}
\end{eqnarray}
Inversion of (\ref{40}-\ref{42}) gives ellipsoidal coordinates in
terms of Cartesian coordinates. Note that ellipsoidal coordinates are only one-to-one in the interior of each octant of the Euclidean space determined by the three coordinate planes $x=0, y=0, z=0$ through the origin.

\subsection{Classical 3D integrable systems.}

Let us study the situation from classical point of view first. Consider a three dimensional system with Hamiltonian given by
$H=\frac{{\bf p}^2}{2m}+U({\bf x})$ and look for two additional
independent integrals of motion of the form:

\begin{equation}\label{43}
\fl
H_1=\frac{1}{2m}p_ig_1^{ij}({\bf x})p_j+U_1({\bf x})\,, \qquad
H_2=\frac{1}{2m}p_ig_2^{ij}({\bf x})p_j+U_2({\bf x})\,,
\end{equation}
with $\{H,H_1\}=\{H,H_2\}=\{H_1,H_2\}=0$, where again the brackets
are the Poisson brackets. The tensors $g_1^{ij}({\bf x})$ and
$g_2^{ij}({\bf x})$ can be obtained using the hypothesis of
commutativity of the kinetic parts of $H$, $H_1$ and $H_2$ (due to
their structure, we could extend the name `Hamiltonian' to $H_i$,
$i=1,2$ too). The condition

\begin{equation}\label{44}
    \{p_i g_1^{ik}({\bf x})p_k,\ {\bf p}^2\}=0
\end{equation}
implies that $p_i g_1^{ik}({\bf x})p_k$ should be a quadratic
function of the momenta. A particular choice follows from a further restriction of $H_1, H_2$ being invariant under any reflection, so that both can be expressed as a
linear combination of the squares $L_i^2$ of the components $L_i$ of the angular momentum
and $p_i^2$ of the linear the momentum $p_i$. There is some arbitrariness in the choice of $H_1$, for which we set

\begin{equation}\label{45}
p_i g_1^{ik}({\bf x})p_k: ={\bf L}^2-p_1^2(b^2+c^2)-p_2^2(a^2+c^2)
-p_3^2 (a^2+b^2)\,,
\end{equation}
where $a$, $b$ and $c$ are the constants which specify the system of
ellipsoidal coordinates, see (\ref{36}).  In matrix form, the tensor
$g_1^{ik}({\bf x})$ defined by (\ref{45}) can be written as

\begin{equation}\label{46}
\fl
g_1^{ik}({\bf x})=\left(
\begin{array}{ccc}
  z^2+y^2-(a^2+c^2) & -xy & -xz \\
  -xy & x^2+z^2-(a^2+c^2) & -yz \\
  -xz & -yz & x^2+y^2-(a^2+b^2) \\
\end{array}
\right)\,.
\end{equation}

Analogously, the commutativity of the kinetic part of $H_2$ with
those of $H$ and $H_1$,

\begin{equation}\label{47}
\{p_i g_2^{ik} ({\bf x}) p_k,\ p_i g_1^{ik}({\bf x}) p_k\}=0
\end{equation}
fixes the linear
combination of the components of ${\bf L}^2$ and of ${\bf p}^2$ in $H_2$ as
\begin{equation}\label{48}
p_i g_2^{ik}({\bf x}) p_k=-
(L_1^2a^2+L_2^2b^2+L_3^2c^2)+p_1^2b^2c^2+p_2^2a^2c^2+p_3^2a^2b^2\,,
\end{equation}
with matrix form given by

\begin{equation}\label{49}
\fl
g_2^{ik}({\bf x})=\left(
\begin{array}{ccc}
  -c^2y^2-b^2z^2+b^2c^2 & c^2xy & b^2xz \\
  c^2xy & -c^2x^2-a^2z^2+a^2c^2 & a^2yz \\
  b^2xz & a^2yz & -b^2x^2-a^2y^2+a^2b^2 \\
\end{array}
\right)\,.
\end{equation}

Matrices (\ref{46}) and (\ref{49}) are symmetric and therefore
diagonalizable. In addition, these matrices commute and therefore
admit a diagonal form in the same basis. This basis is given by

\begin{equation}\label{50}
\fl
E_\lambda:= ( \partial_x\lambda, \partial_y\lambda,
\partial_z\lambda)\,,\;\; E_\mu:=  ( \partial_x\mu, \partial_y\mu,
\partial_z\mu)\,,\;\;E_\nu:= ( \partial_x\nu, \partial_y\nu,
\partial_z\nu)\,.
\end{equation}
These vectors are mutually orthogonal \cite{GaIoNePr07}. Note that
$\lambda$, $\mu$ and $\nu$ are functions of the Cartesian
coordinates, so that the partial derivatives in (\ref{50}) make
sense. The eigenvalues of (\ref{46}) and (\ref{49}) are easy to
obtain and are given in the following table:

\begin{equation}\label{51}
\begin{array}{cccc} \hspace{2cm}   & \qquad g^{ij}({\bf x}) &\qquad
 g^{ij}_1({\bf x}) &\qquad
 g_2^{ij}({\bf x}) \\[2ex]
  E_\lambda & 1 & \mu+\nu & \mu \nu \\
  E_\mu & 1 & \lambda+\nu & \lambda \nu \\
  E_\nu & 1 & \lambda+\mu & \lambda \mu
\end{array}
\end{equation}

Thus, we have determined the kinetic parts of $H_1, H_2$ in terms of ellipsoidal
coordinates. Our next goal is to obtain the most general form of the
`potential' terms $U({\bf x})$, $U_1({\bf x})$ and $U_2({\bf x})$.
From the commutation relations $\{H,H_1\}=\{H,H_2\}=\{H_1,H_2\}=0$,
we obtain the relations,

\begin{equation}\label{52}
\partial^i U_1 =  g_1^{ik}\partial_k U\,,\quad \partial^i U_2 =
 g_2^{ik}\partial_k
U\,, \quad  g_1^{ik}\partial_k U_2 = g_2^{ik}\partial_k  U_1\,.
\end{equation}

Then, after a calculation that makes use of the chain rule involving
ellipsoidal and Cartesian coordinates, the form (\ref{50}) of the
eigenvectors of matrices $g_1^{ik}({\bf x})$ and $g_2^{ik}({\bf x})$
and the orthogonality of these vectors, we can obtain the most
general form of the potentials $U$, $U_1$ and $U_2$ satisfying
$\{H,H_1\}=\{H,H_2\}=\{H_1,H_2\}=0$ \cite{GaIoNePr07}. This is:
\smallskip
\begin{eqnarray}
  U &=& \frac{l(\lambda)}{(\lambda-\nu)(\lambda-\mu)} +
   \frac{m(\mu)}{(\mu-\nu)(\mu-\lambda)} +
   \frac{n(\nu)}{(\nu-\mu)(\nu-\lambda)}\,,\label{53} \\ [2ex]
   U_1 &=& \frac{(\mu+\nu)\, l(\lambda)}{(\lambda-\nu)(\lambda-\mu)} +
   \frac{(\nu+\lambda)\, m(\mu)}{(\mu-\nu)(\mu-\lambda)} +
   \frac{(\lambda+\mu)\, n(\nu)}{(\nu-\mu)(\nu-\lambda)}\,,\label{54}
   \\[2ex]
 U_2 &=& \frac{(\mu\nu)\, l(\lambda)}{(\lambda-\nu)(\lambda-\mu)} +
   \frac{(\nu\lambda)\, m(\mu)}{(\mu-\nu)(\mu-\lambda)} +
   \frac{(\lambda\mu)\, n(\nu)}{(\nu-\mu)(\nu-\lambda)}\,,\label{55}
\end{eqnarray}
where $l(\lambda)$, $m(\mu)$ and $n(\nu)$ are arbitrary functions of
their arguments. This is the most general form of the potentials
compatible with complete integrability.

\subsection{The quantum case.}

We shall briefly comment the procedure here; details can be found in
\cite{GaIoNePr07}. First of all, we obtain the quantum operators $H$,
$H_1$ and $H_2$ by canonical quantization of their classical
counterparts (\ref{43}). Then, complete integrability means that
these three Hamiltonians commute with each other. Therefore, we can
find wave functions $\psi({\bf x})$ such that

\begin{equation}\label{56}
    (H-E)\psi=(H_1-E_1)\psi=(H_2-E_2)\psi=0\,.
\end{equation}
We can prove that the use of ellipsoidal coordinates $\lambda$,
$\mu$ and $\nu$, the form for the potentials given in (\ref{55}) and
the factorization $\psi({\bf x})=\psi(\lambda)\phi(\mu)\varphi(\nu)$
show that equations (\ref{56}) are equivalent to three wave
functions solely in the variables $\lambda$, $\mu$ and $\nu$
respectively, so that complete separation of variables is also
achieved in the quantum case \cite{GaIoNePr07}. The equation for $\lambda$
gives ($\psi_\lambda$ denotes the derivative of $\psi$ with respect to $\lambda$):
\begin{eqnarray}
\fl
4(\lambda+a^2)(\lambda+b^2)(\lambda+c^2)\psi_{\lambda\lambda}
+2[a^2b^2+a^2c^2+b^2c^2
+2\lambda(a^2+b^2+c^2)+3\lambda^2]\psi_\lambda\nonumber \\ \hskip10pt
  +2m (\lambda^2 E-\lambda  E_1+
  E_2-l(\lambda))\psi=0\,.\label{57}
\end{eqnarray}
The other two equations are just obtained by replacing $\lambda$ by
$\mu$ and $\nu$ respectively. Equation (\ref{57}) can be written in
a more compact form by means of an of a change of variable to the
auxiliary variable $t$. This is given by means of the condition:

\begin{equation}\label{58}
    \lambda'(t)=2\sqrt{(\lambda+a^2)(\lambda+b^2)(\lambda+c^2)}\,.
\end{equation}

In terms of $t$, equation (\ref{57}) has the follwing form

\begin{equation}\label{59}
 \frac{d^2\psi}{dt^2}+2m[\lambda^2(t)E-\lambda(t)
    E_1+ E_2-l(\lambda(t))]\psi=0\,,
\end{equation}
where $l(\lambda)$ has already appeared in (\ref{53}-\ref{55}). In
addition to (\ref{59}), there are two other equations, one for $\mu$
and the other for $\nu$. These three equations are similar and, in
particular, all depend on the function $\lambda(t)$ even in the free
particle case, $l(\lambda)=m(\mu)=n(\nu)=0$.

\section{Integrability in spaces of constant curvature.}

There are three types of homogeneous three dimensional manifolds with constant
curvature and Riemannian positive definite metric, which are the sphere $S^3$, the Euclidean plane $E^3$ and the hyperboloid $H^3$ (for general references on integrability in spaces of constant curvature, see \cite{Wozmis03}). In this section, we shall deal with the standard $S^3$ with curvature equal to $1$. As is well known, $S^3$ can be realized as the submanifold $X_1^2 +X_2^2 +X_3^2 +X_4^2 =1$ in a 4D Euclidean ambient space, with the induced metric. A convenient coordinate system for $S^3$ is provided through stereographic coordinates, the point on $S^3$ parametrized by ${\bf x}\in{\mathbb R}^3 \cup \infty$ being given by: 
\begin{equation}\label{60}
    X = ({\bf X}, X_4) = \left( \frac{2\chi {\bf x}}{{\bf x^2}+\chi^2},\quad
\frac{{\bf x^2}-\chi^2}{{\bf x^2}+\chi^2}\right)\,.
\end{equation}
where $\chi$ is a parameter which should be different from zero. With this choice ${\bf x}={\bf 0}$ corresponds to the sphere's `South Pole' $(0,0,0,-1)$, and the projection is made from the North Pole $(0,0,0,1)$ to the plane $X_4=1-\chi$. No actual generality would be lost if we fix a particular value for $\chi$; the preferred choice $\chi=2$ corresponds to projecting over the plane tangent to the sphere's south pole and expressions are generally more clear for this choice, because in this case near the South Pole the coordinates ${\bf x}$ approach the ordinary Cartesian coordinates in the tangent $E^3$ with the same scaling as lengths on the sphere (a rescaling is required for other values of $\chi$) and neglecting terms which are higher order in ${\bf x}$ brings the Euclidean expression with the correct factors, as obvious for instance in (\ref{61})-(\ref{61b}).   

The Lagrangian for the free evolution on $S^3$ can be taken in terms of the ambient space coordinates as:
\begin{equation}\label{61a}
L_0=\frac{m}{2} \left( {\dot
{\bf X}^2} + {\dot{X_4^2}}\right)\,,
\end{equation}
which can be expressed in stereographic coordinates as
\begin{equation}\label{61}
L_0=\frac{m}{2}4\chi^2 \frac{{\bf\dot
x^2}}{({\bf x^2}+\chi^2)^2}\,,
\end{equation}
Alternatively, we may start from the angular momentum tensor in ambient space, with components
$M_{\alpha\beta}:= X_\alpha \dot X_\beta-\dot X_\alpha X_\beta$ and
define the free lagrangian in $S^3$ as (proportional to) the square of this angular momentum tensor; this leads to the same free lagrangian. The canonical conjugate momenta associated to the stereographic coordinates $x_i$ is
\begin{equation}\label{61b}
p_i=m\frac{4\chi^2}{({\bf x^2}+\chi^2)^2}\,\dot x_i.
\end{equation}
and the Legendre transformation provides the following free Hamiltonian:

\begin{equation}\label{62}
H_0=\frac{1}{2m}\frac{1}{4\chi^2}{\bf p^2}({\bf x^2}+\chi^2)^2\,.
\end{equation}

The symmetry group for $S^3$ is $SO(4)$ and (\ref{62}) is invariant
with respect to the sphere isometries in $SO(4)$. Then, $H_0$ has 6
`kinematic' integrals of motion, which are generators of $SO(4)$; in terms of the stereographic parametrization these generators are: 
\begin{equation}\label{63}
L_{i}=\epsilon_{ijk}x_j p_k\,,\qquad K_{i}=\frac{1}{2\chi}\Big(
2 x_i {\bf p}\cdot{\bf x} - p_i({\bf x^2}-\chi^2)
\Big)\,,
\end{equation}
with Poisson brackets 

\begin{equation}\label{64}
\{L_i,L_j\}=-\epsilon_{ijk}L_k\,,\quad
\{K_i,K_j\}=-\epsilon_{ijk}L_k\,, \quad
\{L_i,K_j\}=-\epsilon_{ijk}K_k\,.
\end{equation}
(these close also an $so(4)$ algebra as expected). 
There are two Casimirs given by ${\bf K}^2+{\bf L}^2$ and ${\bf
K}\cdot{\bf L}$. The free Hamiltonian is proportional to the first
Casimir:

\begin{equation}\label{65c}
H_0=\frac1{2m}({\bf K}^2+{\bf L}^2)\,.
\end{equation}
while the second Casimir vanishes. 

Then, we are going to pose in $S^3$ the same question studied for the flat case
in section 1. Assume the existence of a symmetry axis, that,
for the moment, we can identify with the $z$-axis. Then, consider a
Hamiltonian with kinetic term given by $H_0$ plus a time independent
potential $U({\bf x})$,
\begin{equation}\label{66}
H= \frac1{2m}({\bf K}^2+{\bf L}^2) +U({\bf x})\,.
\end{equation}

Then, if this system has to be completely integrable, we need finding a new `Hamiltonian' $H_1$ with `kinetic' part
$T_1=(2m)^{-1}p_ig^{ij}({\bf x})p_j$ and `potential' $\Phi({\bf x})$,
$H_1=T+\Phi({\bf x})$, independent of $H$ and such that
$\{H,L_3\}=\{H_1,L_3\}=\{H,H_1\}=0$. As in Section 1, the kinetic term $T_1$ is obtained under the condition that it commutes with $H_0$ and $L_3$. There is an $S^3$ analog of the most general `kinetic' term (\ref{1a}), but we simply write down the version for $S^3$ of the Euclidean Killing tensor (\ref{1b}) leads to the constant of motion: 
 
\begin{equation}\label{67}
H_1=\frac1{2m}\,(L_1^2+L_2^2- a^2 K_3^2)+\Phi({\bf
x})\,,
\end{equation}
where $a$ is a constant playing a role fully similar to those in (\ref{1b}). If the symmetry axis were arbitrary in
the direction given by the unitary axis ${\bf n}=(n_1,n_2,n_3)$, the
matrix $g^{ij}$ has the following form in terms of the stereographic parametrization:
\begin{eqnarray}
g^{ij}({\bf x})&=&\delta_{ij}({\bf x}\cdot{\bf n})^2-{\bf
x}\cdot{\bf n}(x_i n_j+x_j n_i) +
n_in_j{\bf x}^2\nonumber\\[1ex]
 &-&\frac{a^2}{4\chi^2}\Big({2 x_i}\,{\bf x}\cdot{\bf n}-n_i({\bf
x}^2-\chi^2)\Big)\Big({2 x_j}\,{\bf x}\cdot{\bf n}-n_j({\bf
x}^2-\chi^2)\Big)\,.\label{68}
\end{eqnarray}

The requirement $\{H,H_1\}=0$ leads to the version for $S^3$ of equation
(\ref{6}), which in this case reads

\begin{equation}\label{69}
\partial_i \Phi({\bf x})=G^{ij}({\bf x})\partial_j U({\bf
x})\,, \quad{\rm with}\quad G^{ij}({\bf x})=\frac{g^{ij}({\bf
x})}{({\bf x^2}+\chi^2)^2}\,.
\end{equation}

We shall denote the eigenvalues of $G^{ij}({\bf x})$  by
$\{\mu_+,\mu_-,\mu_3\}$. The two former are the roots of the
quadratic equation $z^2-Pz-Q=0$, where

\begin{equation}\label{70}
    P:= \frac{{\bf x}^2-\alpha^2 [({\bf x}^2-\chi^2)^2+4({\bf
x}\cdot{\bf n})^2]}{({\bf x}^2+\chi^2)^2}\,,\qquad
Q:=\frac{({\bf x}\cdot{\bf n})^2\alpha^2}{({\bf x}^2+1)^2}
\end{equation}
and $\mu_3=Q/\alpha^2$. As in the flat case described in section 1,
the eigenvalues $\mu_\pm$ together with  the azimuthal angle $\varphi$ define
a coordinate system, which is the analogous for the sphere $S^3$ of the spheroidal oblate coordinates \cite{Olevski}. As in the flat case, their corresponding
eigenvectors have components $\partial_i\mu_-$ and
$\partial_i\mu_+$, $i,j=1,2,3$ respectively, i.e.,

\begin{equation}\label{71}
G^{ij}\partial_j\mu_-=\mu_+\partial_i\mu_-\,,\qquad
G^{ij}\partial_j\mu_+=\mu_-\partial_i\mu_+\,.
\end{equation}

Finally, following a procedure similar to that studied in section 1,
we can show that equations (\ref{71}) provide the most general form
for the `potentials' $U({\bf x})$ and $\Phi({\bf x})$ on the variables
$\mu_\pm$. The final result is:
\begin{equation}\label{72}
U=\frac{f(\mu_+)-g(\mu_-)}{\mu_+-\mu_-}\,, \qquad
\Phi=\frac{\mu_-\,f(\mu_+)-\mu_+\,g(\mu_-)}{\mu_+-\mu_-}\,,
\end{equation}
where $f(\mu_+)$ and $g(\mu_-)$ are arbitrary functions. Note that
symmetry prevents that $U$ and $\Phi$ to depend on the azimuthal angle
$\varphi$, as constancy of $L_3$ means invariance under rotations with $z$-axis.

\subsection{Application to the Kepler problem.}

In this subsection, we shall discuss a particular case of very
special importance: the Kepler problem in the standard sphere $S^3$ with curvature equal to 1, which provides  an example of the type of situation just described (recall a factor which dimensionally is a lenght square has been taken as equal to 1, and hence is invisible in the expressions). For more details on the Kepler problem in spaces with constant curvature, see \cite{CRS05,CRS07c,DoZi91,GaSa07,Ko92,NiRoSa99,Pr07}. 

In the sphere $S^3$ (and also in any dimension $n>3$), this system is an example of a maximally superintegrable system. Restricting attention to the 3D case, the Kepler Hamiltonian is
\begin{equation}\label{65}
H=\frac1{2m}({\bf K}^2+{\bf L}^2) - \frac{k}{\tan(r)}
\end{equation}
where the kinetic part is (\ref{66}) and the potential term depends only on the 
intrinsic distance $r$ in $S^3$ to the potential center and $k>0$ for the atractive case. The potential, with center at sphere's south pole, can be expressed in terms of the ambient space coordinates as 
\begin{equation}\label{65a}
- \frac{k}{\tan(r)} = k\frac{X_4}{\sqrt{X_1^2 + X_2^2 + X_3^2}}
\end{equation}
and in terms of stereographic coordinates the Kepler Hamiltonian is: 
\begin{equation}\label{73}
H=\frac1{2m}({\bf K}^2+{\bf L}^2)+{k}\frac{1}{2\chi}\frac{{\bf
x}^2-\chi^2}{|{\bf x}|}\,.
\end{equation}
where the generators in the kinetic part are given in (\ref{63}). 
This Hamiltonian possesses 3 integrals of motion linear in the momenta (the three components of the
angular momentum ${\bf L}$) and a further 3 integrals of motion which are quadratic in the momenta (the components of the Laplace-Runge-Lenz vector $\bf
A$):
\begin{equation}\label{74}
{\bf A}={\bf K}\times{\bf L}+m k \frac{\bf x}{|{\bf x}|},
\end{equation}

Of course there are 2 independent relations among the seven constants of motion $H, {\bf L}, {\bf A}$, reducing to five functionally independent constants, which entitles the Kepler problem in $S^3$ to be maximally superintegrable. These relations are:
\begin{equation}\label{75z}
{\bf A}^2 = m^2 k^2 + \Big( 2m H - {\bf L}^2 \Big) {\bf L}^2, \qquad {\bf A}\cdot{\bf L}=0
\end{equation}

The algebra of Poisson brackets for the components of ${\bf
L}$ and $\bf A$ has the following form:
\begin{equation}\label{75}
\{A_i,A_j\}=\epsilon_{ijk}L_k(2 m H -2{\bf L}^2)\,,\qquad
\{L_i,A_j\}=-\epsilon_{ijk}A_k\,.
\end{equation}

Should these commutation relations be computed in a sphere with curvature $\kappa$, the term $-2{\bf L}^2$ in the first Poisson bracket would appear as
$-2\kappa {\bf L}^2$, making expressions dimensionally correct. This term displays clearly the effects due to the curvature of the configuration space, when it is compared (\ref{75}) with the commutation relations for
the corresponding algebra in the flat case, which are
\begin{equation}\label{75y}
\{A_i,A_j\}=\epsilon_{ijk}L_k 2 m H\,,\qquad
\{L_i,A_j\}=-\epsilon_{ijk}A_k\,.
\end{equation}

From (\ref{75}) we see that the components of ${\bf L}$ and $\bf A$
do not have the commutation relations of a Lie algebra, because of the presence of higher order terms in the Poisson bracket; the quadratic term $L_k H$ is already present in the flat case, but on the curved sphere $S^3$ cubic terms $L_k {\bf L}^2$ appear as well. In the flat Euclidean case, this algebra has been studied under the name of Higgs algebra \cite{GrKu00}. 

Then we consider this system as an example of the situation discussed in the previous section: the Kepler Hamiltonian admits two additional constant of motion which are also in involution: the components of the angular momentum anf of the Runge-Lenz vector along any fixed direction. If we take the $z$ axis, then we get $H, L_3, A_3$ as three constants of motion in involution. 

Now we note that the Kepler problem has axial symmetry around any axis. In particular, the third constant $A_3$ in the Kepler problem in Euclidean  belong indeed to the family (\ref{4}) with the values $\alpha=0, \beta=1, \gamma=0$ for the parameters. While the coordinate system behind the family $\alpha\neq0, \beta=0, \gamma=0$ were oblate spheroidal coordinates in $E^3$ and their analogous in $S^3$, the family $\alpha=0, \beta\neq0, \gamma=0$ turns out to be separable in parabolic coordinates in $E^3$ and in its analogous for $S^3$.  

To end, we mention the possibility, which exists in $S^3$, to rescale the Runge-Lenz vector by a factor which depends on the integrals of motion (so the property of being a constant is not disturbed) in such a way that the Poisson brackets of new components, together with those of angular momentum closes a {\it Lie algebra} $so(4)$. A rescaling ensuring this \cite{Pr07} is:
\begin{equation}\label{76}
{\bf R}={\bf A}\left[{\bf
L}^2 + m \left(\sqrt{ H^2 + k^2} - H \right)\right]^{-1/2}\,,
\end{equation}
(the rescaling factor is always positive, no matter neither the value nor the sign of $H$) and the components of $\bf L$ and $\bf R$ turn out to have the $so(4)$ commutation relations with respect to the Poisson bracket (compare
to (\ref{64})):

\begin{equation}\label{77}
\{L_i,L_j\}=-\epsilon_{ijk}L_k\,,\quad
\{R_i,R_j\}=-\epsilon_{ijk}L_k\,, \quad
\{L_i,R_j\}=-\epsilon_{ijk}R_k\,.
\end{equation}

For these commutation relations (\ref{77}) the first Casimir reads ${\bf
R}^2+{\bf L}^2$. Direct computations reveals that this Casimir is:
\begin{equation}\label{78a}
{\bf R}^2+{\bf L}^2 = m \Big( H + \sqrt{H^2 + k^2}\Big)
\end{equation}
Either by solving this equation for $H$ or by a direct tedious but straightfoward computation, this allows us to write Hamiltonian (\ref{73}) in
terms of this Casimir:

\begin{equation}\label{78}
2 m H ={\bf R}^2+{\bf L}^2-\frac{m^2k^2}{{\bf R}^2+{\bf L}^2}\,.
\end{equation}

This discussion has been purely classical. We simply mention that this last property can be also discussed in the quantum case, where now the new global $so(4)$ Lie algebra symmetry allows us to derive the energy spectrum first found by Schr\"odinger for the Kepler problem in $S^3$ \cite{Sch40,Sch41,St41,IS45}. Full details on this new derivation can be found in \cite{Pr07}; we must recall this approach is conceptually very different from the standard Pauli discussion for the Euclidean Kepler problem, where there is a different Lie algebra in each energy eigenspace, with its isomorphism class depending on the energy sign.

\section{Concluding remarks.}

We have discussed a special class of three dimensional completely integrable either in the flat Euclidean 3D space as well as in the standard sphere $S^3$; these systems are required to have a rotational symmetry axis, so they have axial symmetry. Further to the Hamiltonian $H$, there are two additional constants for such a system: $L$, the angular momentum with respect the symmetry axis, and $H_1=1/(2m)p_ig^{ij}_2({\bf x})p_j+\Psi({\bf x})$. The general form for the tensor $g^{ij}_2({\bf x})$ which guarantees  that the `kinetic' parts of $H$ and $H_2$ commute among themselves and with $L$ is given, and two special cases are mentioned; we discuss one of them, corresponding geometrically to separability in oblate spheroidal coordinates in full detail; the other corresponds to separation of variables in parabolic coordinates.  Integrability determines the `potentials' $V({\bf x}), \Psi({\bf x})$ to have a particular form in terms of undetermined functions of the coordinates, and the
equations of motion are written as three equations each one
involving one coordinate (separation of variables). In the quantum
case, one of the three separated wave equations is trivial and each
of the other two can be easily transformed into a spheroidal wave
equation, for which the solutions have been studied.

If no symmetry conditions are imposed a similar study can be
performed. Separation of variables can be achieved in this case, as one could expect,  in terms of the (general) ellipsoidal coordinates. In the quantum case, we obtain three
similar wave equations, one for each of the variables, each one in
terms of a different function. These functions appear in the form of
the potentials as a consequence of the integrability condition.

Finally, we have carried  the analysis of the systems with axial symmetry to the three dimensional sphere $S^3$ in the classical case. The results obtained are quite similar to the flat ones, in agreement to the idea that results in this area for the constant curvature spaces are essentially `the same' as in the flat case. This is illustrated with the particular case of the
three dimensional Kepler problem.
Extension to higher dimensions should be also possible \cite{DeKuOnVe02,KuTeV01}

\section*{Acknowledgements.}

Partial financial support is acknowledged to the Junta de Castilla y
Le\'on Project VA013C05, the Ministry of Education and Science of
Spain projects MTM2005-09183 and FIS2005-03988 and Grant
SAB2004-0169 and the Russian Science Foundation Grant 04-01-00352.
G.P. was also supported by the programme ENTER-2004/04EP-48,
E.U.-European Social Fund (75\%) and Greek Ministry of development -
GSRT (25\%) and  by RFFI grant 07-01-00234.

\def\otherrefs#1{\null}



\section*{References}
\begin{thebibliography}{10}


\bibitem{BaHeSaSa03}{\sc A.\ Ballesteros, F.J.\ Herranz, M.\ Santander, T.\ Sanz-Gil},
{\rm Maximal superintegrability on $N$-dimensional curved spaces},
{\sl J. Phys. A} {\bf 36}, L93--99 (2003).
\bibitem{Benen97}{\sc S.\ Benenti}, 
{\rm Intrinsic characterization of the variable 
     separation in the Hamilton-Jacobi equation},
{\em J. Math. Phys.} {\bf 38},  6578--6602 (1997).
\bibitem{CRS05}{\sc J.F.\ Cari\~nena,  M.F.\ Ra\~nada, 
                     M.\ Santander}, 
{\rm Central potentials on spaces of constant curvature: The
Kepler problem on the two-dimensional sphere $S^2$ and the
hyperbolic plane $H^2$},
{\sl J. Math. Phys.} {\bf 46}, 052702, 1--25 (2005).
\bibitem{CRS07c}{\sc J.F.\ Cari\~nena,  M.F.\ Ra\~nada, 
                     M.\ Santander}, 
{\rm Superintegrability on curved spaces, orbits and momentum hodographs:  revisiting a classical result by Hamilton},
{\sl J. Phys.A} To appear (2007).
\bibitem{Cram03}{\sc M.\ Crampin}, 
{\rm Conformal Killing tensors with vanishing torsion and the 
     separation of variables in the Hamilton-Jacobi equation}, 
{\em Diff. Geom. Appl.} {\bf 18},  87-102 (2003).
\bibitem{DeKuOnVe02}{\sc B. Demircioglu, \c S. Kuru, M. Onder and A. Vercin},
{\rm  Two families of superintegrable and isospectral potentials in two dimensions},
J. Math. Phys.  {\bf 43}, 2133 (2002).
\bibitem{DoZi91}{\sc P.\ Dombrowski, J.\ Zitterbarth}, 
{\rm On the planetary motion in the 3-Dim standard spaces $M_{\kappa}^3$ of
     constant curvature  $\kappa$},
{\sl  Demonstratio Mathematica} {\bf 24}, 375--458  (1991).
\bibitem{Ev90}{\sc N.W.\ Evans},
{\rm Superintegrability in classical mechanics},
Phys. Rev. {\bf A 41}, 5666--76 (1990).
\bibitem{GaNePr07} {\sc M. Gadella, J. Negro and G.P. Pronko}, J. Phys. A: Math.
Theor., {\bf 40} (2007), 10791.
\bibitem{GaIoNePr07} {\sc M. Gadella, M. Ioffe, J. Negro, G.P. Pronko}, {\rm
Integrable systems in ellipsoidal coordinates}, to appear.
\bibitem{GaSa07}{\sc L. Garc\'{\i}a--Guti\'errez and M.\ Santander}, 
{\rm Levi-Civita regularization and geodesic flows for the `curved' Kepler problem}, 
{\tt arXiv:0707.3810 [math-ph]}
\bibitem{GrKu00}{\sc V.V.\ Gritsev and Yu.A.\ Kurochkin},
The Higgs algebra and the Kepler problem in R3
J. Phys. A  {\bf 33}, 4073--4079, (2000).
\bibitem{HeBa06}{\sc F.J.\ Herranz, A.\ Ballesteros},
{\rm Superintegrability on Three-Dimensional Riemannian and Relativistic Spaces of Constant Curvature},
{\em SIGMA (Symmetry, Integrability and Geometry: Methods and Applications} {\bf  2}, 010, 22p (2006). Available online at {{\tt http://www.emis.de/journals/SIGMA/2006/  Paper010/}}
\bibitem{HoMcLSm05}{\sc J.T.\ Horwood, R.G.\ McLenaghan, R.G.\ Smirnov},
{\rm  Invariant classification of orthogonally separable Hamiltonian systems in Euclidean space},
Comm. Math. Phys.  {\bf 259}, (2005) 679--709.
\bibitem{IS45}{\sc L. Infeld and A. Schild},  
{\rm A note on the Kepler problem in a space of constant 
     negative curvature},
Phys. Rev. {\bf 67}, 121--122 (1945).
\bibitem{Ka99}{\sc E.G. Kalnins, G.C. Williams, W. Miller, G.S.  Pogosyan},
{\rm Superintegrability in three-dimensional Euclidean space},
J. Math. Phys.  {\bf 40},  (1999)  708-725.
\bibitem{Ka02}{\sc E.G. Kalnins, J.M. Kress, W. Miller, G.S. Pogosyan},
{\rm Complete sets of invariants for dynamical systems that admit a separation of variables},
J. Math. Phys.  {\bf 43},  (2002)  708-725.

\bibitem{Ko92}{\sc V.V. Kozlov and A.O. Harin},
{\rm Kepler's problem in constant curvature spaces},
Celest. Mechanics  {\bf 54},  (1992)  393--399.
\bibitem{KuTeV01}{\sc \c S. Kuru, A. Te\u gmen and A. Vercin},
{\rm  Intertwined isospectral potentials in an arbitrary dimension},
J. Math. Phys.  {\bf 42},  (2001)   3344--3360.
\bibitem{Wi67}{\sc A. A. Makarov, J. A. Smorodinsky, K. Valiev, and P. Winternitz},
{\rm  A systematic search for nonrelativistic systems with dynamical symmetries},
Nuovo Cim. A  {\bf 52},  (1967)  1061-1084.
\bibitem{MeScWoBKMaFu} {\sc J. Meixner,
F.W. Sch\"afke, and G. Wolf}, {\it Mathieu Functions and Spheroidal
Functions and Their Mathematical Foundations}, Springer-Verlag,
1980.
\bibitem{MiPaZa81}{\sc W. Miller, J. Patera and P. Winternitz},
{\rm  Subgroups of Lie groups and separatio of variables},
J. Math. Phys.  {\bf 22}, (1981) 251--260.
\bibitem{NiRoSa99}{\sc L.M. Nieto, H.C. Rosu and M. Santander},
{\rm  Hydrogen atom as an eigenvalue problem in 3D spaces of constant curvature and minimal length},
Mod. Phys. Lett. A  {\bf 14}
, 2463--2469, (1999).
\bibitem{Olevski}{\sc M.N.\ Olevski}, 
{\rm Triorthogonal systems in spaces of constant curvature 
     in which the equation  $\Delta_3u +\lambda u=0$ allows a 
     complete separation of variables}, 
{\em Mat. Sb. {\bf 27} (69)}, 379--426 (1950) (In Russian). 
\bibitem{Pr07} {\sc G.P. Pronko}, {\rm Kepler problem in the constant
curvature space}, {\tt arXiv:0705.3111v2 [math-ph]} 
\bibitem{SaSa05}{\sc M.\ Santander and T.\ Sanz-Gil}, 
{\rm From oscillator(s) and Kepler(s) potentials to general superintegrable
     systems in spaces of constant curvature}, 
{\sl Rep. on Math. Phys.}, {\bf 55}, pp. 371--383, (2005). 
\bibitem{Sch40}{\sc E.\ Schr\"odinger},
{\rm A method of determining quantum mechanical eigenvalues and 
     eigenfunctions},
Proc. R.I.A. {\bf A 46}, 9--16 (1940).
\bibitem{Sch41}{\sc E. Schr\"odinger},
{\rm  Further studies on solving eigenvalue problems by factorization},
Proc. Roy. Irish Acad. {\bf A 46},  (1941) 183--206.
\bibitem{St41}{\sc A.F.\ Stevenson},
{\rm Note on the `Kepler Problem' in a spherical space, and the 
     factorization method of solving eigenvalue problems},
Phys. Rev. {\bf 59}, 842--843 (1941).
\bibitem{Wozmis03}{\sc T.G.\ Vozmischeva},
{\it Integrable problems of celestial mechanics in spaces of 
constant curvature},
Astrophysics and Space Science Library, 295.
(Kluwer Academic Pub., Dordrecht, 2003).

\end{thebibliography}
\end{document}